\documentclass[12pt,]{article}
\usepackage{authblk}
\usepackage{fullpage}
\usepackage{amssymb,amsmath}
\usepackage[utf8x]{inputenc}
\usepackage[T1]{fontenc}
\usepackage{siunitx}
\usepackage[version=3]{mhchem}
\usepackage{graphicx}
\usepackage{algorithm}
\usepackage{algpseudocode}

\usepackage[left]{lineno}
\newcommand{\indep}{\perp \!\!\! \perp}

\usepackage{setspace}
\title{Fourier Diffusion Models: A Method to Control MTF and NPS in Score-Based Stochastic Image Generation}
\author[1]{\small Matthew Tivnan}
\author[1]{Jacopo Teneggi}
\author[2]{Tzu-Cheng Lee}
\author[2]{Ruoqiao Zhang}
\author[3]{Kirsten Boedeker}
\author[2]{\\Liang Cai}
\author[4]{Grace J. Gang}
\author[1]{Jeremias Sulam}
\author[1]{J. Webster Stayman}

\affil[1]{\scriptsize Department of Biomedical Engineering, Johns Hopkins University, Baltimore, MD, USA}
\affil[2]{Canon Medical Research, USA. Vernon Hills, IL, USA}
\affil[3]{Canon Medical Systems Corporation, Otawara, Japan}
\affil[4]{Department of Radiology, Hospital of the University of Pennsylvania, Philadelphia, PA, USA}
\date{\small \today}

\begin{document}

\maketitle

% \newpage

\begin{abstract}
    Score-based stochastic denoising models have recently been demonstrated as powerful machine learning based tools for conditional and unconditional image generation. The existing methods are based on a forward stochastic process wherein the training images are scaled to zero over time and white noise is gradually added such that the final time step is approximately zero-mean identity-covariance Gaussian noise. A neural network is then trained to approximate the time-dependent score function, or the gradient of the logarithm of the probability density function, for that time step. Using this score estimator, it is possible to run an approximation of the time-reversed stochastic process to sample new images from the training data distribution. These score-based generative models have been shown to out-perform generative adversarial neural networks using standard benchmarks and metrics. However, one issue with this approach is that it requires a large number of forward passes of the neural network. Additionally, the images at intermediate time steps are not directly useful, since the signal-to-noise ratio is low. In this work we present a new method called Fourier Diffusion Models which replaces the scalar operations of the forward process with shift-invariant convolutions and the additive white noise with additive stationary noise. This allows for control of MTF and NPS at intermediate time steps. Additionally, the forward process can be crafted to converge to the same MTF and NPS as the measured images. This way, we can model continuous probability flow from true images to measurements. In this way, the sample time can be used to control the tradeoffs between measurement uncertainty and generative uncertainty of posterior estimates. We compare Fourier diffusion models to existing scalar diffusion models and show that they achieve a higher level of performance and allow for a smaller number of time steps. 
\end{abstract}

\newpage
\section{Introduction}

Denoising diffusion probabilistic models \cite{sohl2015deep, ho2020denoising} and closely-related score-based generative modeling through stochastic differential equations \cite{song2020score} have recently been demonstrated as powerful machine learning based tools for conditional and unconditional image generation. These diffusion models are based on a stochastic process in which the true images are degraded over time through deterministic decay of the signal and additive random noise. Then, a neural network is trained to approximate the time-reversed stochastic process starting with low-quality images from a known prior distribution (pure noise in many cases), iteratively running reverse time update steps, and eventually ending on an approximate sample from the same distribution as the training images. Compared to another popular method, generative adversarial neural networks, diffusion models can achieve higher image quality as measured with standard benchmarks while avoiding the difficulties of adversarial training \cite{dhariwal2021diffusion}.

In this article, we present a new method called \emph{Fourier Diffusion Models,} which allow for  control of MTF and NPS at each time step of the forward and reverse stochastic process. Our approach is to model the forward process as a cascade of LSI systems with ASGN. Then, we train a neural network to approximate the time-dependent score function for iterative sharpening and denoising of the images to generate high-quality posterior samples given measured images with spatial blur and textured noise. The new feature of Fourier diffusion models compared to conventional scalar diffusion models is the capability to model continuous probability flow from ground truth images to measured images with a certain MTF and NPS, rather than ending the forward process on pure noise. We hypothesize that Fourier diffusion models will require fewer time steps for conditional image generation relative to scalar diffusion models because the true images are more similar to measured images than they are to white noise.

In the sections to follow, we provide detailed mathematical descriptions of Fourier diffusion models including the forward stochastic process, the training loss function for score-matching neural networks, and instructions on how to sample the reverse process for conditional or unconditional image generation. Finally, we present experimental methods and results for image restoration of low-radiation-dose CT measurements to demonstrate one practical application of the proposed method.

\section{Methods}

\subsection{Linear Shift-Invariant Systems with Stationary Gaussian Noise}

In this section, we describe the theoretical background for LSI systems with ASGN. This is a standard mathematical model used to evaluate the spatial resolution and noise covariance of medical imaging systems in the spatial frequency domain.

% While this approximation may not be exact for physical systems, it is a useful approximation and often used as a standard for image quality evaluation.  

% \subsubsection{Linear Shift-Invariant Systems}

A two-dimensional LSI system is mathematically defined by convolution with the impulse response function, also known as the point spread function (PSF), of the system. Fourier convolution theorem states that it is equivalent to multiply the two-dimensional Fourier transform of the input by the Fourier transform of the PSF, referred to as the modulation transfer function (MTF) of the system, followed by the inverse Fourier transform to produce the convolved output. For discrete systems, the voxelized values of a medical image can be represented as a flattened column vector and the convolution operation can be represented by a circulant matrix operator, $\mathbf{H}_\text{MTF} = \mathbf{U}^*_\text{DFT} \boldsymbol{\Lambda}_\text{MTF} \mathbf{U}_\text{DFT}$, where $\mathbf{U}_\text{DFT}$ is the unitary discrete two-dimensional Fourier transform, $\mathbf{U}^*_\text{DFT}$ is the unitary discrete two-dimensional inverse Fourier transform, and  $\boldsymbol{\Lambda}_\text{MTF}$ is a diagonal matrix representing element-wise multiplication by the MTF in the spatial frequency domain. 

% \subsubsection{LSI Systems with Additive Stationary Gaussian Noise}

If we assume the noise covariance between two voxels in an image does not depend on position, only on relative displacement between two positions, then we say the noise is stationary, and the covariance can be fully defined by the noise power spectrum (NPS) in the spatial frequency domain. For the discrete case, stationary noise can be modeled by a circulant covariance matrix, $\boldsymbol{\Sigma}_\text{NPS} = \mathbf{U}^*_\text{DFT} \boldsymbol{\Lambda}_\text{NPS} \mathbf{U}_\text{DFT}$, where $\boldsymbol{\Lambda}_\text{NPS}$ is a diagonal matrix of spatial-frequency-dependent noise power spectral densities. In probabilistic terms, the image quality can be modeled by a multivariate Gaussian conditional probability density function parameterized by the MTF and NPS as follows:

\begin{equation}
    p(\mathbf{x}_\text{out}|\mathbf{x}_\text{in}) =  \mathcal{N}(\mathbf{x}_\text{out};  [\mathbf{U}^*_\text{DFT} \boldsymbol{\Lambda}_\text{MTF} \mathbf{U}_\text{DFT}] \hspace{1mm} \mathbf{x}_\text{in}, [\mathbf{U}^*_\text{DFT} \boldsymbol{\Lambda}_\text{NPS} \mathbf{U}_\text{DFT}] ).
\end{equation}

% This model describes spatial resolution and noise covariance in the spatial frequency domain. This is particularly useful for evaluating the detectability of image features of a certain size.

% \subsubsection{MTF and NPS in Cascaded LSI Systems}

% In some cases it is useful to consider a sequence or cascade of operations where each stage consists of one LSI system, $\mathbf{H}^{(n)}$ and additive Gaussian stationary noise with NPS, . We can show that  

\subsection{Discrete-Time Stochastic Process with MTF and NPS Control}

To train score-based diffusion models for medical image restoration, we begin by defining a forward stochastic process in which the image quality is degraded over time. Then, the goal is to train a neural network to run an approximation of the time-reversed stochastic process. That is, the model is trained to sample high-quality images at earlier time steps given low-quality images at later time steps. In this section, we describe a method to control the evolution of MTF and NPS in the discrete-time forward stochastic process to allow for spatial-frequency-dependent control of spatial resolution and noise covariance.

\begin{figure}
    \centering
    \includegraphics[width=\textwidth]{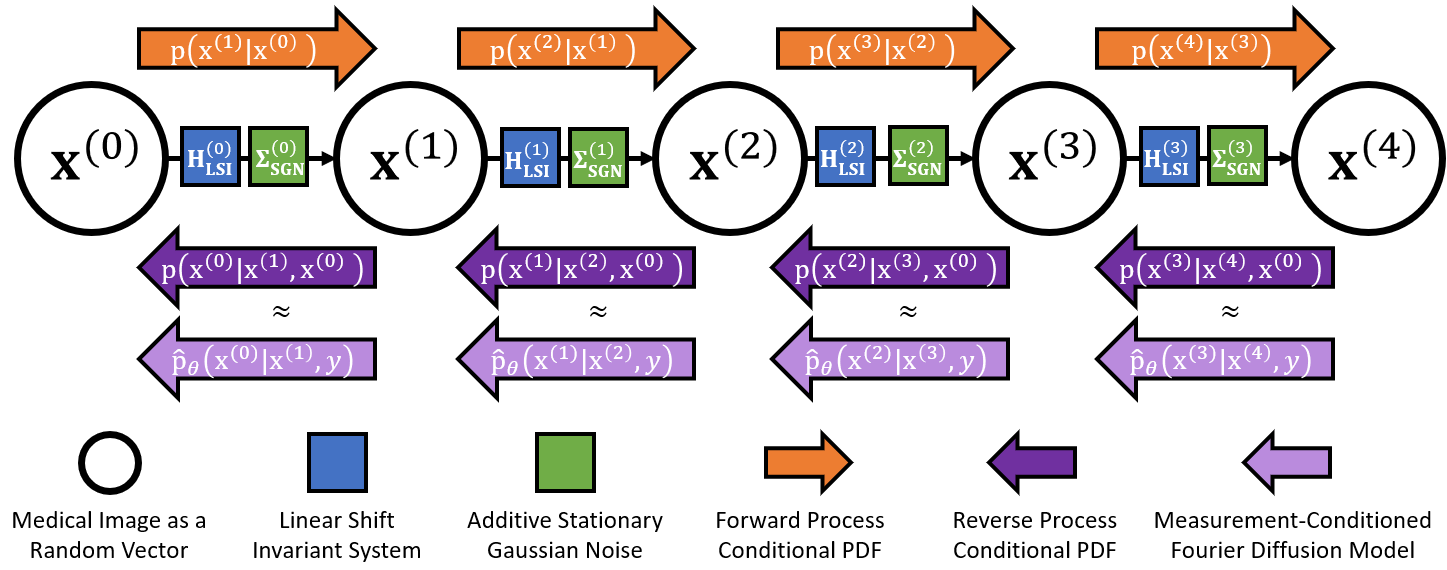}
    \caption{A probabilistic graphical model for the stochastic process, consisting of linear shift invariant systems and additive stationary Gaussian noise. The  measurement-conditioned diffusion model, in light purple, is trained to approximate the reverse process, in dark purple.}
    \label{fig:network_diagram}
\end{figure}

% \subsubsection{Discrete-Time Forward Process: Cascaded LSI Systems with ASGN}

Consider a sequence of LSI systems with ASGN resulting in a discrete-time forward stochastic process with the following update rule:

\begin{equation}
\mathbf{x}^{[n+1]} = \mathbf{H}_\text{LSI}^{[n]} \mathbf{x}^{[n]} + \boldsymbol{\Sigma}_\text{SGN}^{1/2}\boldsymbol{\eta}^{[n]},
\end{equation} 

\noindent where $\mathbf{H}_\text{LSI}^{[n]}$ is a circulant matrix representing an LSI system, $\boldsymbol{\Sigma}_\text{SGN}$ is a circulant matrix representing the noise covariance of the ASGN,  and $\boldsymbol{\eta}^{[n]}$  is zero-mean identity-covariance Gaussian noise. Also, we assume the noise at a given time step is independent of the noise at all other time steps; that is, $\boldsymbol{\eta}^{[n]} \indep \boldsymbol{\eta}^{[m]} \hspace{1mm} \forall \hspace{1mm} n \neq m$. Therefore, the conditional probability density function for a forward step is

\begin{equation}
    \text{p}(\mathbf{x}^{[n+1]}|\mathbf{x}^{[n]}) = \mathcal{N}(\mathbf{x}^{[n+1]} ; \mathbf{H}_\text{LSI}^{[n]} \hspace{1mm} \mathbf{x}^{[n]},  \boldsymbol{\Sigma}_\text{SGN}) .
    \label{eq:forward_step}
\end{equation}

As shown in Figure \ref{fig:network_diagram}, the probabilistic graphical model for this forward process  is a directed acyclic graph. Stated differently, the image at a given time step, $\mathbf{x}^{[n]}$, depends only on the previous image, $\mathbf{x}^{[n-1]}$. Therefore, the joint  distribution of the full forward stochastic process can be written as

\begin{equation}
\text{p}(\mathbf{x}^{[0]}, \mathbf{x}^{[1]},\mathbf{x}^{[2]},\ldots,\mathbf{x}^{[N]}) = \text{p}(\mathbf{x}^{[0]})\text{p}(\mathbf{x}^{[1]}| \mathbf{x}^{[0]})\text{p}(\mathbf{x}^{[2]}| \mathbf{x}^{[1]})\ldots\text{p}(\mathbf{x}^{[N]}| \mathbf{x}^{[N-1]}) \enspace ,
\end{equation}

\noindent where $N$ is the last time step. 

One example would be the case where $\mathbf{H}_\text{LSI}^{[n]}$ is a small amount of convolutional blur. In that case, the image will become more blurry and noisy as time passes in the forward process. Then, as we will describe in a later section, a neural network can be trained to run the time-reversed stochastic process, which should result in a series of sharpening and noise-reducing steps to restore image quality.

% \subsubsection{Prescriptive Control of MTF and NPS in the Forward Process}

 The MTF and NPS of the image at a given time in the forward process can be described by an equivalent LSI system, $\mathbf{H}_\text{MTF}^{[n]} =  \mathbf{U}^*_\text{DFT} \boldsymbol{\Lambda}_{\text{MTF}}^{[n]} \mathbf{U}_\text{DFT}$, and ASGN with covariance, $\boldsymbol{\Sigma_\text{NPS}^{[n]}}=\mathbf{U}^*_\text{DFT}\boldsymbol{\Lambda}_{\text{NPS}}^{[n]}\mathbf{U}_\text{DFT}$, applied to the original image, $\mathbf{x^{[0]}}$ as shown below:

\begin{gather}
    \mathbf{x}^{[n]} = \mathbf{H}_\text{MTF}^{[n]} \mathbf{x}^{[0]} + {\boldsymbol{\Sigma}_\text{NPS}^{[n]}}^{ \hspace{-2mm} 1/2} \boldsymbol{\epsilon}^{[n]}\\
    \text{p}(\mathbf{x}^{[n]}|\mathbf{x}^{[0]}) = \mathcal{N}(\mathbf{x}^{[n]} ;   \mathbf{H}_\text{MTF}^{[n]} \hspace{1mm} \mathbf{x}^{[0]},  \boldsymbol{\Sigma_\text{NPS}^{[n]}} ) .
    \label{eq:effective_MTF_NPS}
\end{gather}

\noindent where $\boldsymbol{\epsilon}^{[n]}$ is identity-covariance zero-mean Gaussian noise. Our goal is to prescribe the effective MTF and NPS at every time step, and then define the forward process parameters accordingly. To that end, we can combine \eqref{eq:forward_step} and \eqref{eq:effective_MTF_NPS} to define $\mathbf{H}_\text{LSI}^{[n]}$ and $\boldsymbol{\Sigma}_\text{SGN}^{[n]}$ in terms of the target MTF and NPS as follows:

\begin{gather}
    \mathbf{H}_\text{LSI}^{[n]}  = \mathbf{H}^{[n+1]}_\text{MTF} \mathbf{H}^{-1 \hspace{0.5mm} [n]}_\text{MTF} = \mathbf{U}^*_\text{DFT} \boldsymbol{\Lambda}_{\text{MTF}}^{[n+1]} \boldsymbol{\Lambda}_{\text{MTF}}^{-1 \hspace{0.5mm} [n]} \mathbf{U}_\text{DFT} \label{eq:LSI_in_terms_of_MTF_}\\
    \boldsymbol{\Sigma}_\text{SGN}^{[n]} = \boldsymbol{\Sigma}_{\text{NPS}}^{[n+1]} - \mathbf{H}_\text{LSI}^{[n]} \boldsymbol{\Sigma}_{\text{NPS}}^{[n]}  \mathbf{H}_\text{LSI}^{[n]} =  \mathbf{U}^*_\text{DFT}[ \boldsymbol{\Lambda}_{\text{NPS}}^{[n+1]} - \boldsymbol{\Lambda}_{\text{NPS}}^{[n]} \boldsymbol{\Lambda}_{\text{MTF}}^{2 \hspace{0.5mm} [n+1]} \boldsymbol{\Lambda}_{\text{MTF}}^{-2 \hspace{0.5mm} [n]}] \mathbf{U}_\text{DFT} \label{eq:noise_in_terms_of_NPS}
\end{gather}

\subsection{Continuous-Time Process and Stochastic Differential Equations}

The forward stochastic process in \eqref{eq:effective_MTF_NPS} can be viewed as discrete time samples of a continuous-time stochastic process, $\mathbf{x}^{(t)}$, given by:

\begin{gather}
\mathbf{x}^{(t)} = \mathbf{H}^{(t)}_\text{MTF} \mathbf{x}^{(0)} + {\boldsymbol{\Sigma}_\text{NPS}^{(t)}}^{\hspace{-3mm}  1/2} \boldsymbol{\epsilon}^{(t)}  
\label{eq:x_t} \\
\text{p}(\mathbf{x}^{(t)}|\mathbf{x}^{(0)}) = \mathcal{N}(\mathbf{x}^{(t)}; \mathbf{H}_\text{MTF}^{(t)} \mathbf{x}^{(0)}, {\boldsymbol{\Sigma}_\text{NPS}^{(t)}}) \label{eq:x_t_dist}
\end{gather}

\noindent where $\boldsymbol{\epsilon}^{(t)}$ is a zero-mean identity-covariance Gaussian process where the updates for non-overlapping time intervals are independent (i.e., a L\'{e}vy process). In Appendix A, we show that this process can be described by the following stochastic differential equation:

\begin{equation}
    \mathbf{dx} = \mathbf{H^{'}}^{(t)}_\text{MTF}{\mathbf{H}^{(t)}_\text{MTF}}^{\hspace{-3mm}-1} \mathbf{x}^{(t)} \text{dt} + (-  2 {\mathbf{H^{'}}^{(t)}_\text{MTF}} {\mathbf{H}^{(t)}_\text{MTF}}^{\hspace{-3mm}-1}  \boldsymbol{\Sigma}_\text{NPS}^{(t)} + \boldsymbol{\Sigma^{'}}_\text{NPS}^{(t)}  )^{1/2} \mathbf{dw} \label{eq:SDE}
\end{equation}

\noindent where $\mathbf{H^{'}}^{(t)}_\text{MTF} = \frac{\text{d}}{\text{dt}} \mathbf{H}^{(t)}_\text{MTF} $,  $\boldsymbol{\Sigma^{'}}_\text{NPS}^{(t)} = \frac{\text{d}}{\text{dt}} \boldsymbol{\Sigma}_\text{NPS}^{(t)}$, and $\mathbf{dw}$ is infinitesimal white Gaussian noise with covariance, $\text{dt} \mathbf{I}$ (as seen in Brownian motion for example). If we compare $\eqref{eq:SDE}$ to the standard form, 

\begin{equation}
    \mathbf{dx} = \mathbf{f}(\mathbf{x}, t) \text{dt} + \mathbf{g}(t) \mathbf{dw},
\end{equation}

\noindent then we can identify,

\begin{gather}
    \mathbf{f}(\mathbf{x}, t) = \mathbf{H^{'}}^{(t)}_\text{MTF}{\mathbf{H}^{(t)}_\text{MTF}}^{\hspace{-3mm}-1} \mathbf{x}^{(t)} \hspace{2mm},\hspace{4mm} 
    \mathbf{g}(t) = (-  2 {\mathbf{H^{'}}^{(t)}_\text{MTF}} {\mathbf{H}^{(t)}_\text{MTF}}^{\hspace{-3mm}-1}  \boldsymbol{\Sigma}_\text{NPS}^{(t)} + \boldsymbol{\Sigma^{'}}_\text{NPS}^{(t)}  )^{1/2} .
    \label{eq:f_g}
\end{gather}

It has previously been shown there is an exact solution for the time-reversed stochastic differential equation \cite{anderson1982reverse}. The formula is shown below:

\begin{equation}
    \mathbf{dx} = [\mathbf{f}(\mathbf{x}, t) - \mathbf{g}^2(t) \nabla_{\mathbf{x}^{(t)}} \log{\text{p} (\mathbf{x}^{(t)})}] \text{dt} + \mathbf{g}(t) \mathbf{dw}.
\end{equation}

\noindent The inclusion of the score function, $\nabla_{\mathbf{x}^{(t)}} \log{\text{p} (\mathbf{x}^{(t)})}$, results in a deterministic drift towards higher probability values of $\mathbf{x}^{(t)}$. For the case where we also have measurements, $\mathbf{y}$, to guide the image generation, the conditional reverse process is:

\begin{equation}
    \mathbf{dx} = [\mathbf{f}(\mathbf{x}, t) - \mathbf{g}^2(t) \nabla_{\mathbf{x}^{(t)}} \log{\text{p} (\mathbf{x}^{(t)}|\mathbf{y})}] \text{dt} + \mathbf{g}(t) \mathbf{dw}.
    \label{eq:reverse_sde_standard_form}
\end{equation}

\noindent We will continue to use the formulae for measurement-conditioned image generation with the understanding that the unconditional case can be obtained by substituting the empty set, $\mathbf{y}=\emptyset$, resulting in $\log{\text{p} (\mathbf{x}^{(t)}|\mathbf{y})} = \log{\text{p} (\mathbf{x}^{(t)})}$.  Substituting the values in \eqref{eq:f_g} into \eqref{eq:reverse_sde_standard_form} results in the following formula for the reverse stochastic differential equation:

\begin{gather}
     \mathbf{dx} = [\mathbf{H^{'}}^{(t)}_\text{MTF}{\mathbf{H}^{(t)}_\text{MTF}}^{\hspace{-3mm}-1} \mathbf{x}^{(t)} - (-  2 {\mathbf{H^{'}}^{(t)}_\text{MTF}} {\mathbf{H}^{(t)}_\text{MTF}}^{\hspace{-3mm}-1}  \boldsymbol{\Sigma}_\text{NPS}^{(t)} + \boldsymbol{\Sigma^{'}}_\text{NPS}^{(t)} )  \nabla_{\mathbf{x}^{(t)}} \log{\text{p} (\mathbf{x}^{(t)}|\mathbf{y})}] \text{dt} \nonumber \\
    \hspace{70mm} + (-  2 {\mathbf{H^{'}}^{(t)}_\text{MTF}} {\mathbf{H}^{(t)}_\text{MTF}}^{\hspace{-3mm}-1}  \boldsymbol{\Sigma}_\text{NPS}^{(t)} + \boldsymbol{\Sigma^{'}}_\text{NPS}^{(t)}  )^{1/2} \mathbf{dw} .
\end{gather}

\subsection{Score-Matching Loss Function for Neural Network Training }

\noindent Recent work \cite{song2020score}, has demonstrated stochastic image generation is possible by training a neural network to approximate the score function $\mathbf{s}_{\boldsymbol{\theta}}(\mathbf{x}^{(t)}, \mathbf{y}, t) \approx \nabla_{\mathbf{x}^{(t)}} \log{\text{p} (\mathbf{x}^{(t)}|\mathbf{y}})$ given  many examples of the training data $\{\mathbf{x}^{(0)}, \mathbf{y}\}$. Then, the score estimator can be used to run an approximation of the reverse stochastic process. In this section, we closely follow the derivations in that previous work with some modifications for Fourier diffusion models. 

We assume that, $\mathbf{x}^{(t)}$, is conditionally independent of the measurements, $\mathbf{y}$, given the true images, $\mathbf{x}^{(0)}$, 
which means, $\nabla_{\mathbf{x}^{(t)}} \log{\text{p} (\mathbf{x}^{(t)}|\mathbf{y}}, \mathbf{x}^{(0)}) = \nabla_{\mathbf{x}^{(t)}} \log{\text{p}(\mathbf{x}^{(t)}|\mathbf{x}^{(0)})}$. Therefore, the score-matching loss function used for neural network training is the Fisher divergence between the model and data distributions, as shown below:

\begin{equation}
    \underset{\mathbf{x}^{(0)}, t}{\mathbb{E}}[||\mathbf{s}_{\boldsymbol{\theta}}(\mathbf{x}^{(t)}, \mathbf{y}, t) - \nabla_{\mathbf{x}^{(t)}} \log{\text{p} (\mathbf{x}^{(t)}|\mathbf{x}^{(0)})}||^2] .
\end{equation}

\noindent for the Gaussian conditional distribution defined in \eqref{eq:x_t_dist}, the score can be substituted as:

\begin{gather}
    \underset{\mathbf{x}^{(0)}, t}{\mathbb{E}}[||\mathbf{s}_{\boldsymbol{\theta}}(\mathbf{x}^{(t)}, \mathbf{y}, t) - {\boldsymbol{\Sigma}_{\text{NPS}}^{(t)}}^{\hspace{-2mm}-1} (\mathbf{x}^{(t)} - \mathbf{H}^{(t)}_{\text{MTF}} \mathbf{x}^{(0)}) ||^2]\\
    \underset{\mathbf{x}^{(0)}, t}{\mathbb{E}}[||\mathbf{s}_{\boldsymbol{\theta}}(\mathbf{x}^{(t)}, \mathbf{y}, t) - {\boldsymbol{\Sigma}_{\text{NPS}}^{(t)}}^{\hspace{-2mm}-1/2} \boldsymbol{\epsilon}^{(t)} ||^2] 
\end{gather}

\noindent Using the parameterization, $\mathbf{s}_{\boldsymbol{\theta}}(\mathbf{x}^{(t)}, \mathbf{y}, t) = {\boldsymbol{\Sigma}_{\text{NPS}}^{(t)}}^{\hspace{-2mm}-1/2} \boldsymbol{\hat{\epsilon}}_{\boldsymbol{\theta}}(\mathbf{x}^{(t)}, \mathbf{y}, t)$, leads to the following:

\begin{equation}
\underset{\mathbf{x}^{(0)}, t}{\mathbb{E}}[(\boldsymbol{\hat{\epsilon}}_{\boldsymbol{\theta}}(\mathbf{x}^{(t)}, \mathbf{y}, t) - \boldsymbol{\epsilon}^{(t)})^T {\boldsymbol{\Sigma}_{\text{NPS}}^{(t)}}^{\hspace{-2mm}-1} (\boldsymbol{\hat{\epsilon}}_{\boldsymbol{\theta}}(\mathbf{x}^{(t)}, \mathbf{y}, t) - \boldsymbol{\epsilon}^{(t)}) ] \label{eq:score_matching_loss}
\end{equation}

After the score-matching neural network is trained, one can run a discrete-time approximation of the reverse process using the Euler-Maryuama method, as shown below:

\begin{gather}
     \mathbf{x}^{[n-1]} = \mathbf{x}^{[n]} - [\mathbf{H^{'}}^{(t)}_\text{MTF}{\mathbf{H}^{(t)}_\text{MTF}}^{\hspace{-3mm}-1} \mathbf{x}^{[n]} - (-  2 {\mathbf{H^{'}}^{(t)}_\text{MTF}} {\mathbf{H}^{(t)}_\text{MTF}}^{\hspace{-3mm}-1}  \boldsymbol{\Sigma}_\text{NPS}^{(t)} + \boldsymbol{\Sigma^{'}}_\text{NPS}^{(t)} )  \mathbf{s}_{\boldsymbol{\theta}}(\mathbf{x}^{[n]}, \mathbf{y}, t)] \Delta t \nonumber \\
    \hspace{70mm} + (-  2 {\mathbf{H^{'}}^{(t)}_\text{MTF}} {\mathbf{H}^{(t)}_\text{MTF}}^{\hspace{-3mm}-1}  \boldsymbol{\Sigma}_\text{NPS}^{(t)} + \boldsymbol{\Sigma^{'}}_\text{NPS}^{(t)}  )^{1/2} \sqrt{\Delta t} \boldsymbol{\zeta}^{[n]} \label{eq:euler-maryuama}
\end{gather}

\noindent where $\boldsymbol{\zeta}^{[n]}$ is zero-mean identity-covariance Gaussian noise where different time steps are independent.

In \cite{song2020score}, the updates of the forward stochastic process are defined by scalar multiplication and additive white Gaussian noise. The contribution of this work is to generalize the definition of the update steps to allow for linear shift invariant systems and additive stationary Gaussian noise. For the special case of the scalar multiplicative drift and white noise, which we refer to as scalar diffusion models, one should set $\mathbf{H}_\text{MTF}^{(t)}$ and $\boldsymbol{\Sigma}_\text{NPS}^{(t)}$ to scalar matrices follows:

\begin{gather}
    \mathbf{H}_\text{MTF}^{(t)} = e^{- \frac{1}{2}\int_0^t \beta(s)\text{ds}}\hspace{1mm} \mathbf{I} \label{eq:scalar_drift}\\
    \boldsymbol{\Sigma}_\text{NPS}^{(t)} = \sigma^2(t) \hspace{1mm} \mathbf{I} \label{eq:white_noise}
\end{gather}

\noindent which results in the forward stochastic differential equation:

\begin{gather}
    \mathbf{dx} = -\frac{1}{2} \beta^{(t)} \mathbf{x}^{(t)}\text{dt} + \sqrt{\beta(t)\sigma^2(t) + \frac{\text{d}}{\text{dt}}\sigma^2(t)}\mathbf{dw} ,
\end{gather}

\noindent where $\sigma^2(t)$ controls the so-called variance-exploding component and $\beta(t)$ controls the variance-preserving component. For the special case of the original denoising diffusion probabilistic models \cite{sohl2015deep} \cite{ho2020denoising}, there is the additional constraint, $\sigma^2(t) = 1 - e^{-\int_0^{t}\beta(s)\text{ds}}$, meaning the process is fully defined by $\beta(t)$. 

One way to interpret Fourier diffusion models is to consider them as conventional diffusion models in the spatial frequency domain. For example, it would be mathematically equivalent to take the two-dimensional Fourier transform of the training images, and then train a conventional diffusion model to generate new samples of those Fourier coefficients using spatial-frequency-dependent diffusion rates (diagonal matrices). One practical advantage of formulating the process with LSI systems is that the reverse time steps may be more suitable for approximation with convolutional neural networks, which are also composed of shift-invariant operators in the image domain.

%  While we were originally motivated by the goal of controlling MTF and NPS in the forward process, the derivations in the previous sections have not relied on any special properties of circulant matrices.  Therefore, we can write another more general version of these equations that applies any Gaussian stochastic process composed of linear systems and additive Gaussian noise without specifying shift invariant systems or stationary noise. We show the general version of this stochastic differential equations, score-matching loss function, and reverse process sampling procedure in Appendix B.  For the experiments in the following section, we focus on the forward process in \eqref{eq:x_t} and score-matching loss function in \eqref{eq:score_matching_loss} using cirulant matrices to train measurement-conditioned Fourier diffusion models for medical image restoration. In the future, we are interested in exploring other applications of the more general equations above.

\subsection{Experimental Methods: Low-Dose CT Image Restoration}

To implement Fourier diffusion models, we need a training dataset of ground truth images as well as a definition of the  MTF and NPS as a function of time in the forward process. In this section, we describe an implementation of our proposed method for low-dose thoracic CT image restoration. 
%We also describe a method to parameterize the MTF and NPS with a short list of coefficients for spatial frequency bands. 
% Finally, we provide an algorithmic description of each step in the training process. 

% \subsubsection{The LIDC Training Dataset: Low-Radiation-Dose Thoracic CT Images }

We used the publicly available Lung Image Database Consortium (LIDC) dataset, which consists of three-dimensional image volumes reconstructed from thoracic CT scans for lung imaging. The first 80\% of image volumes were used for training data and the last 20\% were reserved for validation data. We randomly extracted 8000 two-dimensional axial slices from the training volumes and 2000 axial slices from the validation volumes. The slices were registered to a common coordinate system using bilinear interpolation so that all images are $512\times512$ with $1.0$~mm voxel spacing. The image values were shifted and scaled such that 0.0 represents -1000~HU and 10.0 represents 1000~HU. These images were used as the ground-truth for training and validation; however, it is important to note that these images still contain errors such as noise, blur, and artifacts which may impact the trained diffusion models. Our approach is to simulate lower-quality images that one may measure with a low-radiation-dose CT scan by applying convolutional blur and adding stationary noise. These low-quality CT images represent the output of a low-dose CT scan, so they are treated as the measurements for the purposes of this study. Our goal is to train conditional score-based diffusion models to sample posterior estimate images given low-dose CT measurements. If successful, the posterior estimate images sampled by the trained model should have similar image quality to the normal-dose training images in the LIDC dataset. 

% \subsubsection{Parameterizing MTF and NPS Trajectories with Band-Pass Filters}

Our proposed approach involves prescribing the MTF and NPS at each stage in the forward process. For this implementation, we chose to parameterize both MTF and NPS using a set of band-pass filters.  Let the matrix $\boldsymbol{\mathcal{G}}(h) =  \mathbf{U}^*_\text{DFT} \boldsymbol{\Lambda}_{\boldsymbol{\mathcal{G}}}(h) \mathbf{U}_\text{DFT}$ represent an isotropic Gaussian low-pass filter in the two-dimensional spatial frequency domain, where $h$ describes the PSF width in units of $\text{mm}$.  That is, the diagonal of $\boldsymbol{\Lambda}_{\boldsymbol{\mathcal{G}}}(h)$ is the Fourier transform of a convolutional Gaussian blur kernel proportional to $\exp{(-\frac{1}{2} (\sqrt{x^2 + y^2})^2 / h^2)}$.  We used the following parameterization of low-pass, band-pass, and high-pass filters:

\begin{gather}
    \mathbf{H}_\text{LPF} = \boldsymbol{\mathcal{G}}(3.0~\text{mm}) \\
    \mathbf{H}_\text{BPF} = \boldsymbol{\mathcal{G}}(1.0~\text{mm}) - \boldsymbol{\mathcal{G}}(3.0~\text{mm}) \\
    \mathbf{H}_\text{HPF} = \mathbf{I} - \boldsymbol{\mathcal{G}}(1.0~\text{mm}) 
\end{gather}

\noindent Our model of low-dose CT measurements, $\mathbf{y}$, given the ground truth image, $\mathbf{x}$ is

\begin{gather}
    \text{p}(\mathbf{y}|\mathbf{x}) = \mathcal{N}(\mathbf{y}; \mathbf{H}_{\mathbf{y}|\mathbf{x}} \hspace{0.5mm} \mathbf{x}, \boldsymbol{\Sigma}_{\mathbf{y}|\mathbf{x}})\\
    \mathbf{H}_{\mathbf{y}|\mathbf{x}} = (1.0) \enspace \mathbf{H}_\text{LPF}  + (0.5) \enspace \mathbf{H}_\text{BPF}  + (0.1) \enspace \mathbf{H}_\text{HPF} \\ 
    \boldsymbol{\Sigma}_{\mathbf{y}|\mathbf{x}} = (0.1) \enspace \mathbf{H}_\text{LPF}  + (1.0) \enspace \mathbf{H}_\text{BPF}  + (0.5) \enspace \mathbf{H}_\text{HPF}  
\end{gather}

\noindent This particular choice of measured MTF and NPS is arbitrary but intended to roughly match the typical patterns observed in low-dose CT images. In general, one can substitute the MTF and NPS to match the calibrated values for a medical imaging system.

In this experiment, we compare two cases: 1) scalar diffusion models using multiplicative drift and additive white Gaussian noise and 2) Fourier diffusion models using linear shift invariant systems and additive stationary Gaussian noise. As shown in \eqref{eq:scalar_drift} and \eqref{eq:white_noise}, the scalar diffusion models are a special case of Fourier diffusion models that can be written using scalar matrices. We define the scalar diffusion model using the following parameters:

\begin{gather}
    \mathbf{H}_\text{MTF}^{(t)} = (e^{-  
 5t^2})\hspace{1mm} \mathbf{I} \label{eq:scalar_drift}\\
    \boldsymbol{\Sigma}_\text{NPS}^{(t)} = (1 - e^{- 10 t^2}) \hspace{1mm} \mathbf{I} \label{eq:white_noise}
\end{gather}

\noindent This forward stochastic process begins with $\mathbf{x}^{(t=0)}$ having the same distribution as the true images, $ \mathbf{x}$, and converges to approximately zero-mean identify-covariance Gaussian noise at the final time step $\mathbf{x}^{(t=1)}$. For the Fourier diffusion model case, we design the forward stochastic process so that the final time step has the same distribution as the low-dose CT measurements; that is, $\mathbf{x}^{(t=1)} \approx \mathbf{y}$. This capability to model continuous probability flow from true images to measured images is a new feature of Fourier diffusion models. For this case, we use the formulae shown below:

\begin{gather}
    \mathbf{H}_\text{MTF}^{(t)} = (1.0) \enspace \mathbf{H}_\text{LPF}  + (0.5 + 0.5e^{-5t^2}) \enspace \mathbf{H}_\text{BPF}  + (0.1 + 0.9e^{-5t^2}) \enspace \mathbf{H}_\text{HPF} \\ 
    \boldsymbol{\Sigma}_\text{NPS}^{(t)} = (0.1 - 0.1 e^{- 10 t^2}) \enspace \mathbf{H}_\text{LPF}  + (1.0 - 1.0 e^{- 10 t^2}) \enspace \mathbf{H}_\text{BPF}  + (0.5 - 0.5 e^{- 10 t^2}) \enspace \mathbf{H}_\text{HPF}  
\end{gather}

\begin{figure}
    \centering
    \includegraphics[width=0.95\textwidth]{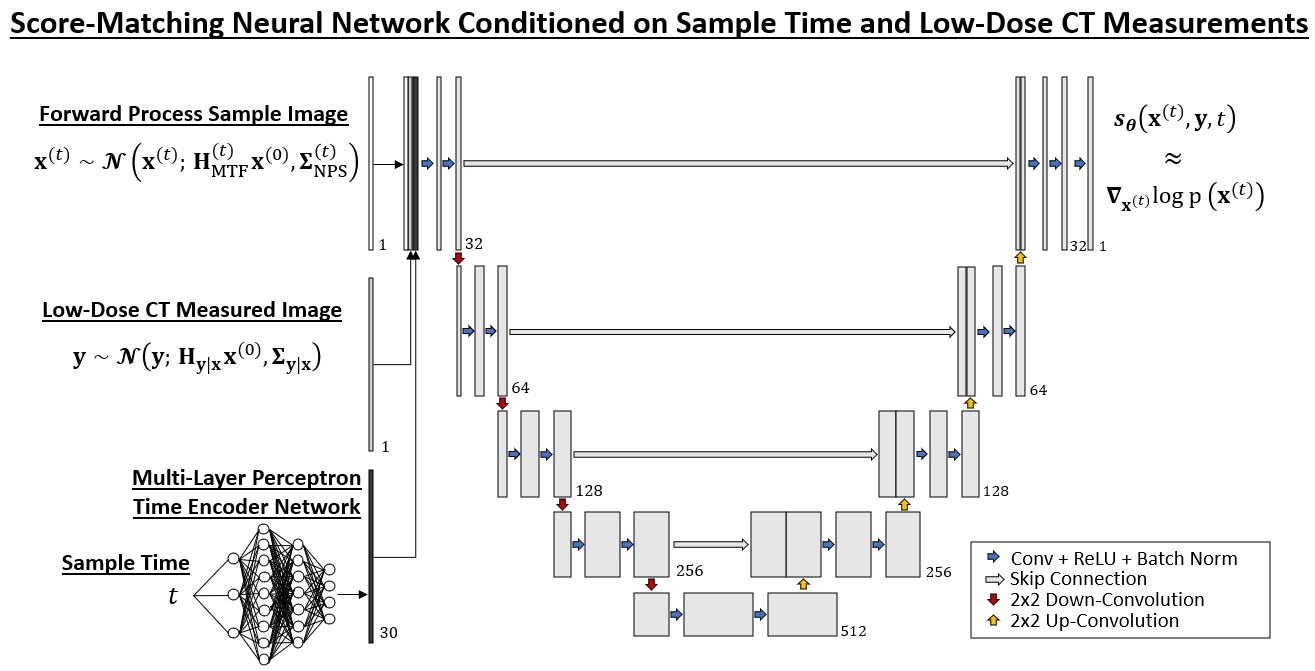}
    \caption{Diagram of the score-matching neural network. The inputs are the forward process sample image, low-dose CT measured image, and sample time. The output is the estimated score function.  }
    \label{fig:unet_diagram}
\end{figure}

\begin{figure}[ht!]
    \centering
\includegraphics[width=0.48\textwidth]{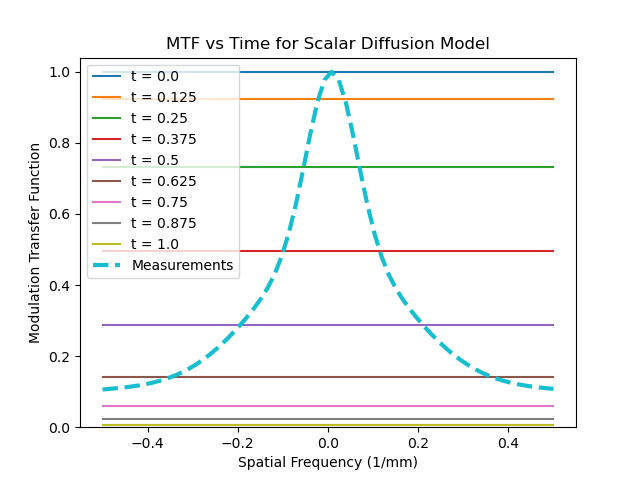}
\includegraphics[width=0.48\textwidth]{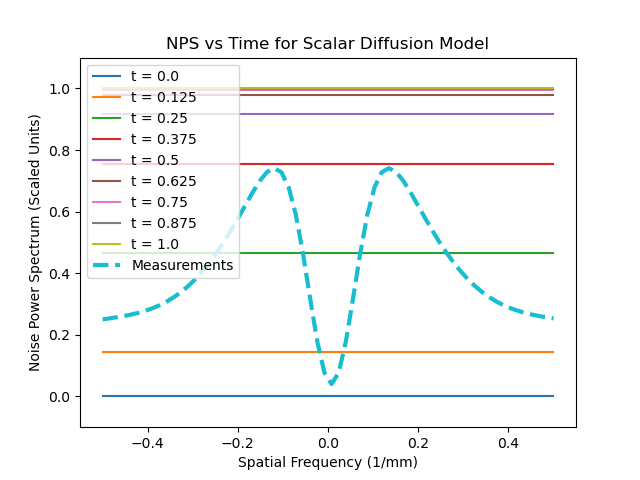}
    \caption{MTF  and NPS vs time for the scalar diffusion model. }
    \label{fig:MTF_NPS_vs_time_scalar}
\end{figure}

\begin{figure}[ht!]
    \centering
\includegraphics[width=0.48\textwidth]{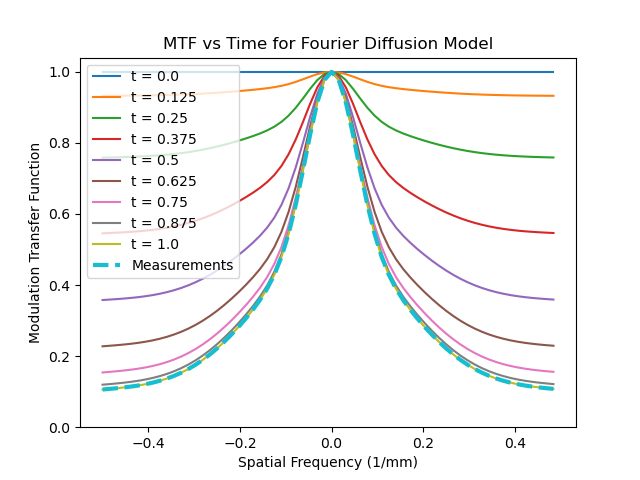}
\includegraphics[width=0.48\textwidth]{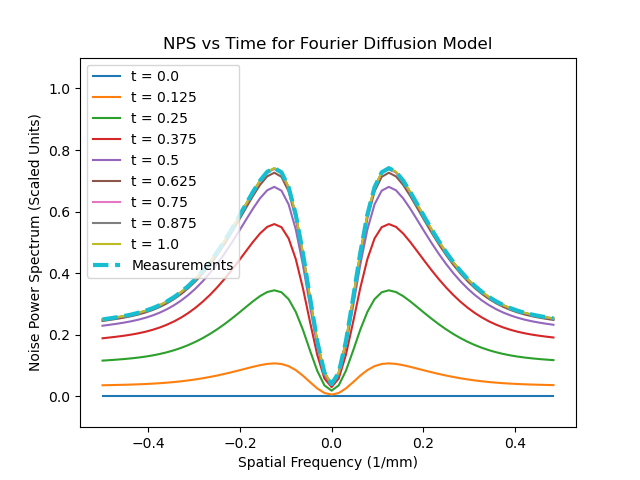}
    \caption{MTF and NPS vs time for the Fourier diffusion model. }
    \label{fig:MTF_NPS_vs_time_fourier}
\end{figure}

For the score-matching neural network, we used the u-net architecture shown in Figure \ref{fig:unet_diagram}. The model inputs are the forward process sample, the low-dose CT measurements, and the sample time. The model output is an estimation of the score function. For time encoding, we applied a multi-layer perceptron to the sample time and converted the output to constant-valued images. The forward process sample image, the low-dose CT measured image, and the time encoding images are concatenated and passed to the score-matching u-net. Each convolutional block consists of a convolutional layer, rectified linear units activation, and batch normalization. The final output layer has no activation function (linear) or batch normalization. Dropout layers were also applied to each convolutional block with a drop out rate of 20\%.  We used the Adam optimizer with a learning rate of $10^{-3}$ \cite{kingma2014adam}. All machine learning modules were implemented with in Pytorch \cite{NEURIPS2019_9015}. We ran 10,000 training epochs, 32 images per batch, and the training loss function in \eqref{eq:score_matching_loss}, where the expectation over $\mathbf{x}^{(0)}$ is implemented via the sample mean over multiple training images per batch and the expectation over time, $t$, is implemented by sampling a different time step independently for each image so that there are also 32 time samples per batch.

After training, we run the reverse process using \eqref{eq:euler-maryuama} for both diffusion models. We discretize the reverse process with 1024, 512, 256, 128, 64, 32, 16, 8 and 4 time steps uniformly spaced between $t=0$ and $t=1$, inclusively. That way, we can analyze the error due to time discretization for scalar and Fourier diffusion models. We ran the reverse process 32 times using the same measurements. For the posterior estimates at $t=0$ of the reverse process, we compute the mean squared error, mean squared bias, and mean variance where the mean refers to a spatial average over the image, error/bias are with respect to the ground truth, $\mathbf{x}$, and the variance refers to the ensemble of samples from the reverse process.

\begin{figure}[ht!]
    \centering
\includegraphics[trim={50mm 20mm 50mm 20mm},clip, width=0.99\textwidth]{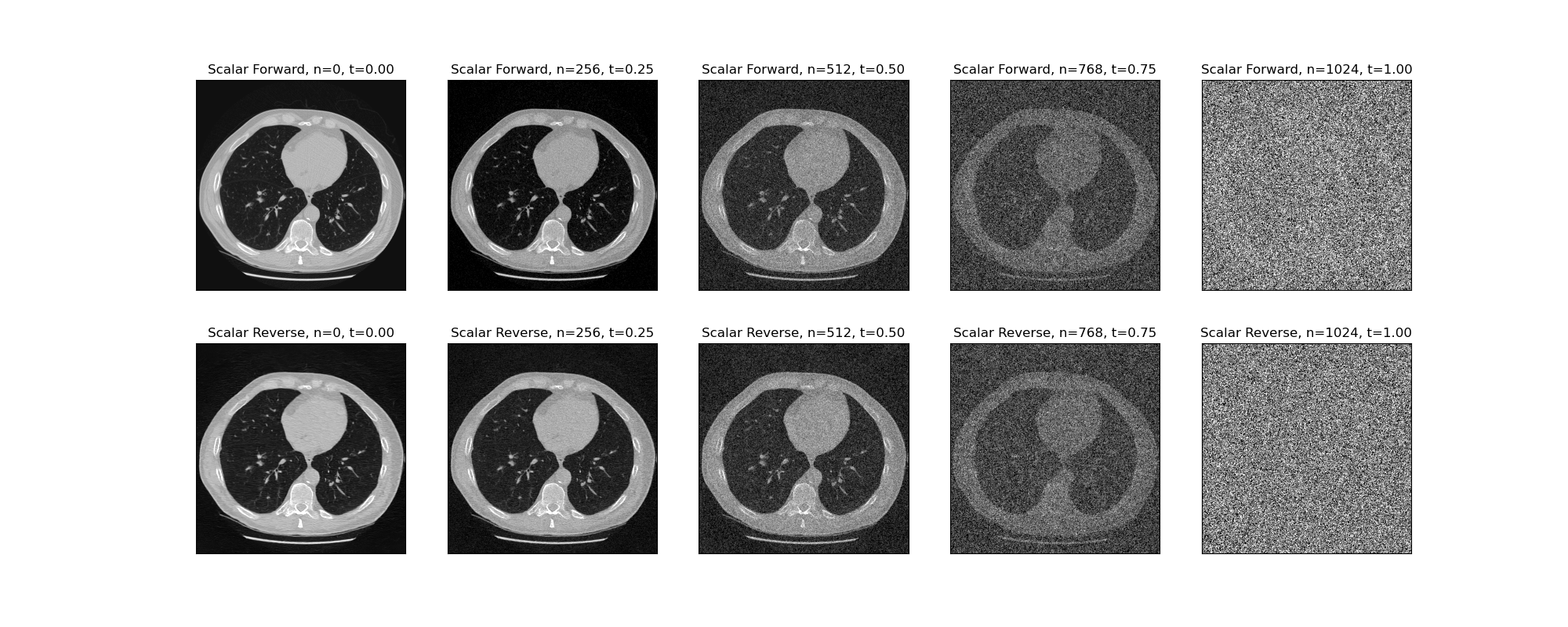}
    \caption{Forward and reverse stochastic process for the scalar diffusion model with 1024 time steps applied to a full CT image.  }
    \label{fig:process_scalar_1024}
\end{figure}

\begin{figure}[ht!]
    \centering
\includegraphics[trim={50mm 20mm 50mm 20mm},clip,width=0.99\textwidth]{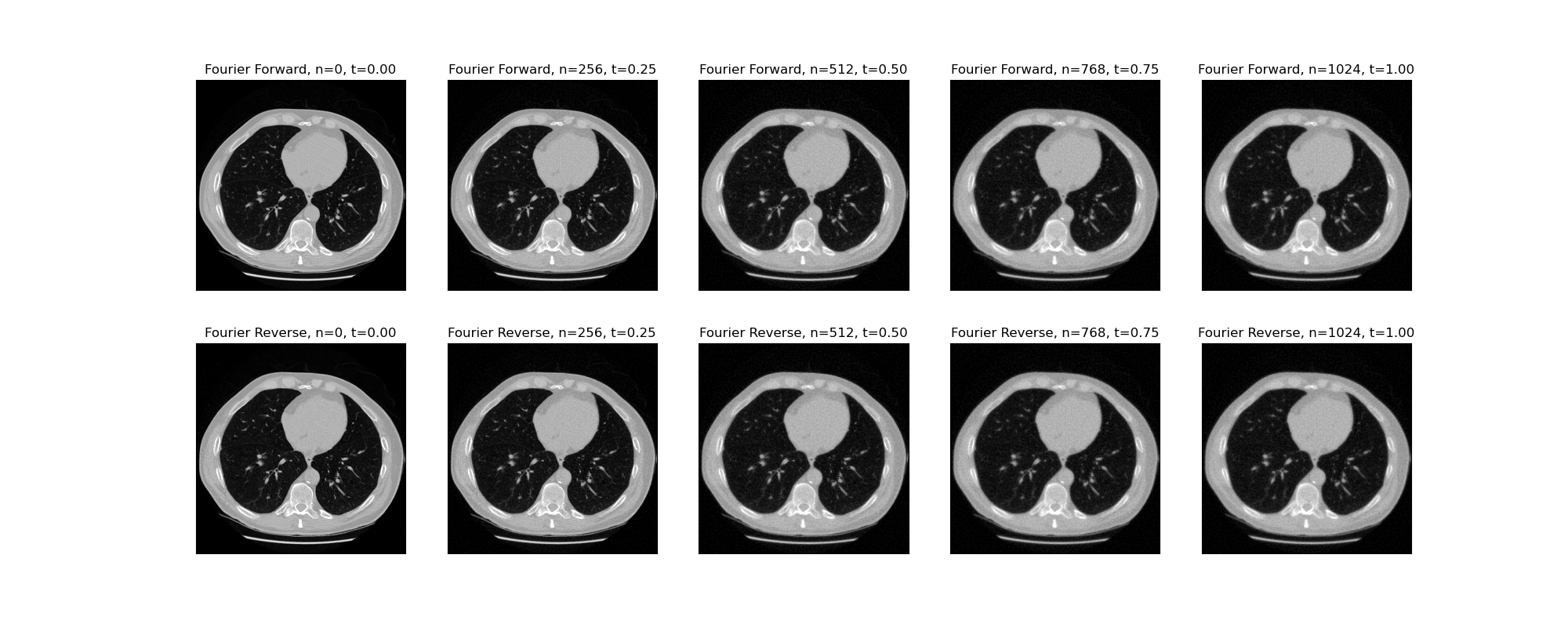}
    \caption{Forward and reverse stochastic process for the Fourier diffusion model with 1024 time steps applied to a full CT image.  }
    \label{fig:process_fourier_1024}
\end{figure}

\begin{figure}[ht!]
    \centering
\includegraphics[trim={50mm 20mm 50mm 20mm},clip,width=0.99\textwidth]{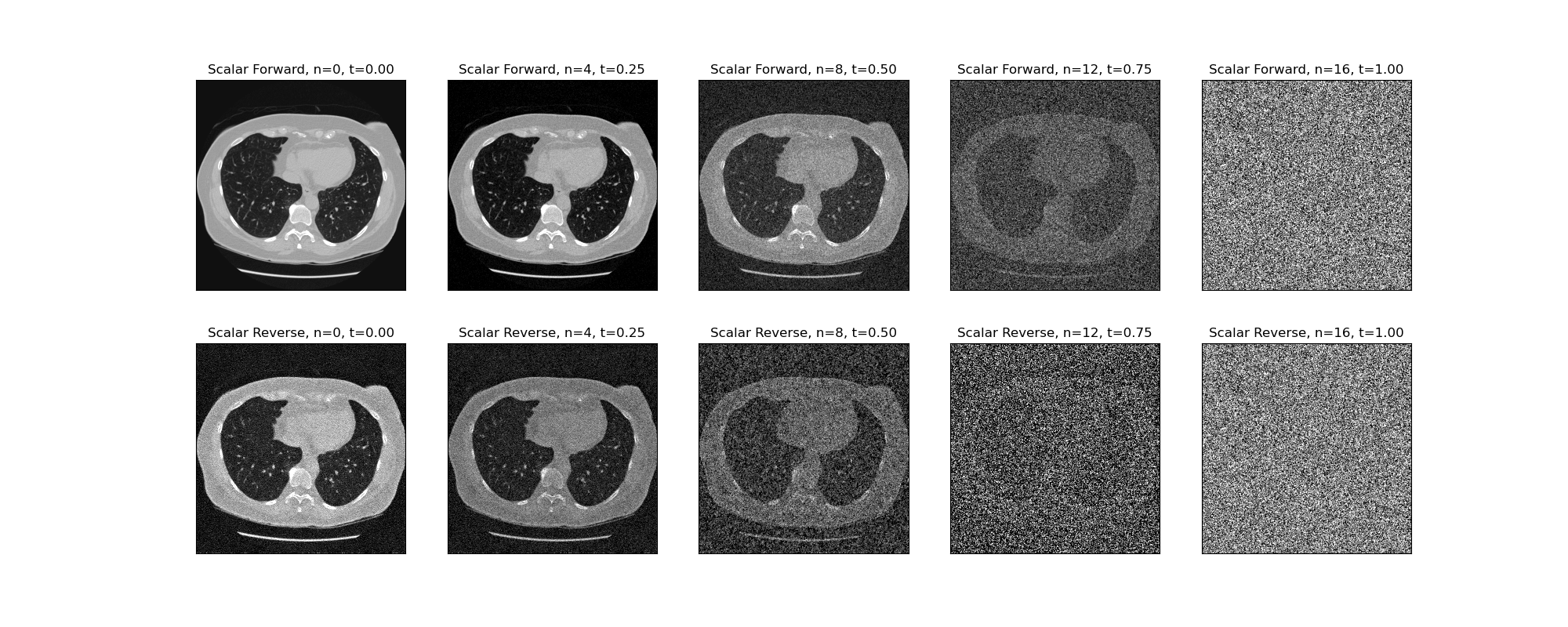}
    \caption{Forward and reverse stochastic process for the scalar diffusion model with 16 time steps applied to a full CT image.  }
    \label{fig:process_scalar_16}
\end{figure}

\begin{figure}[ht!]
    \centering
\includegraphics[trim={50mm 20mm 50mm 20mm},clip,width=0.99\textwidth]{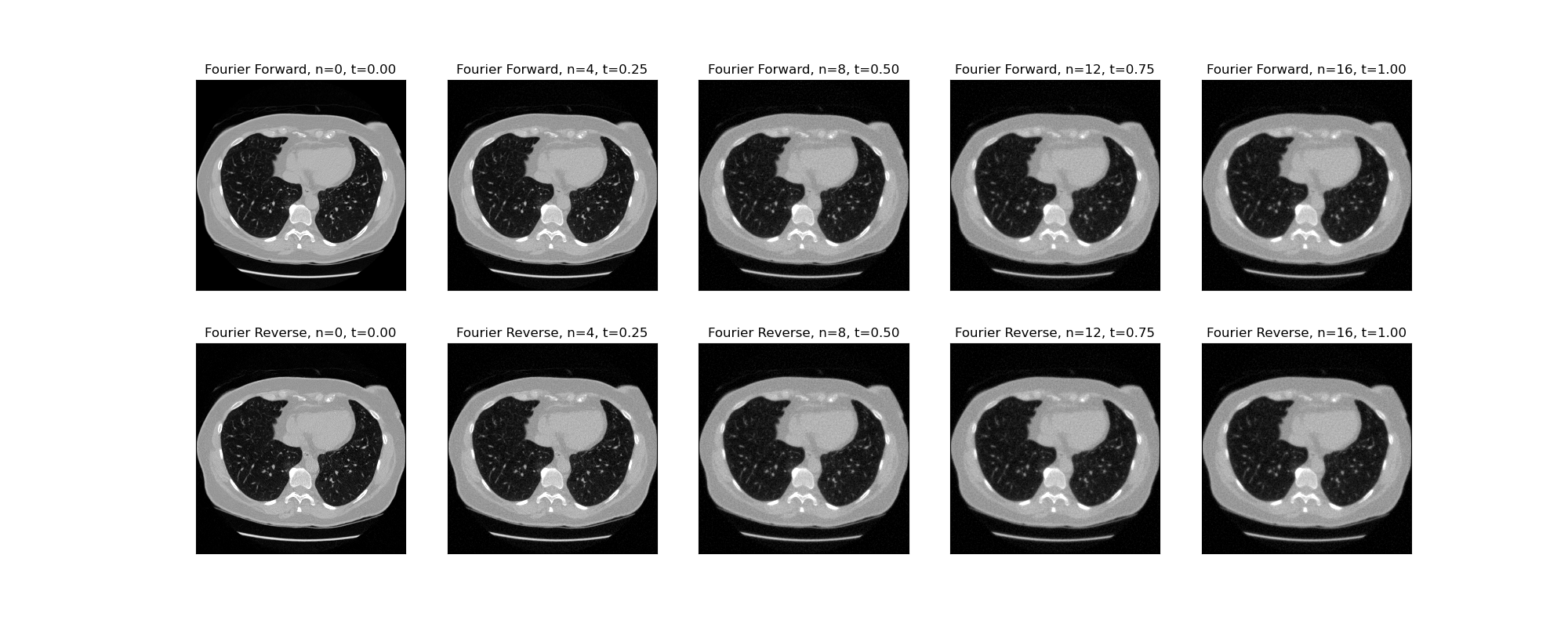}
    \caption{Forward and reverse stochastic process for the Fourier diffusion model with 16 time steps applied to a full CT image. }
    \label{fig:process_fourier_16}
\end{figure}

\begin{figure}[ht!]
    \centering
\includegraphics[trim={50mm 20mm 50mm 20mm},clip,width=0.99\textwidth]{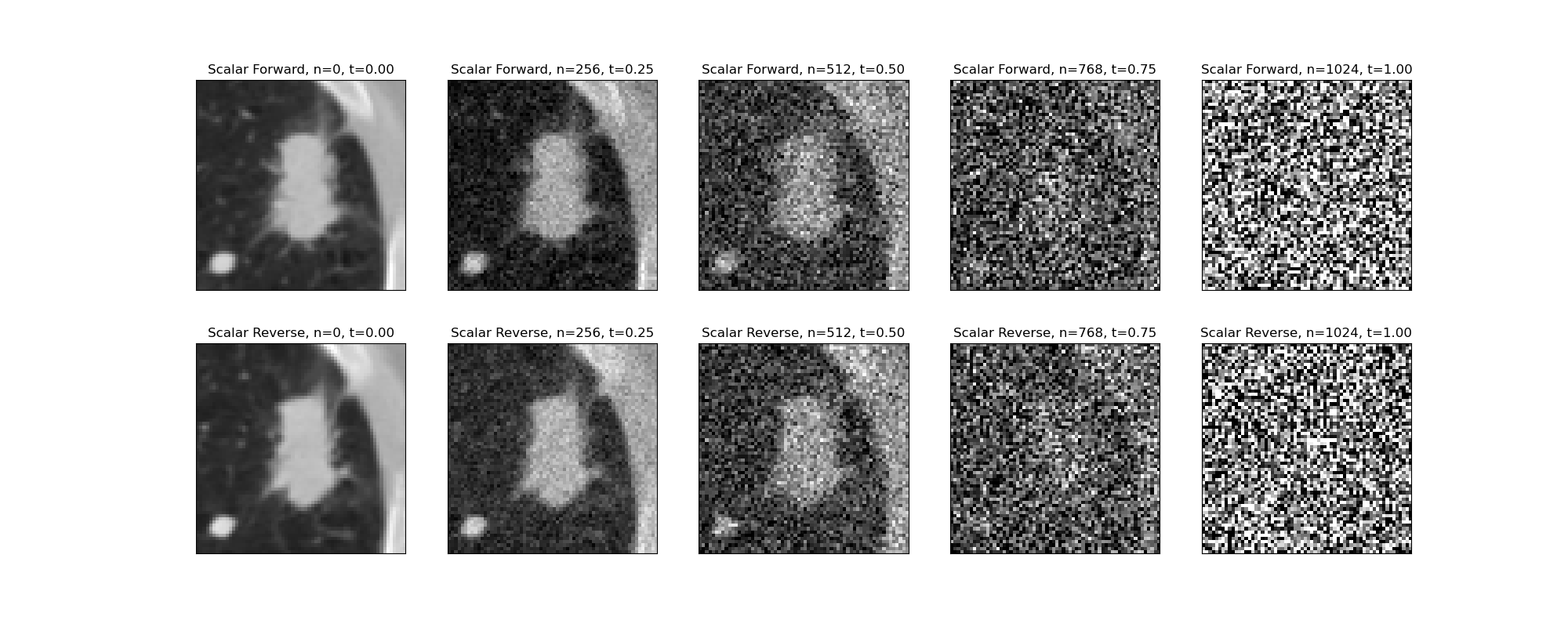}
    \caption{Forward and reverse stochastic process for the scalar diffusion model with 1024 time steps applied to an image patch showing a lung nodule.  }
    \label{fig:process_patch_scalar_1024}
\end{figure}

\begin{figure}[ht!]
    \centering
\includegraphics[trim={50mm 20mm 50mm 20mm},clip,width=0.99\textwidth]{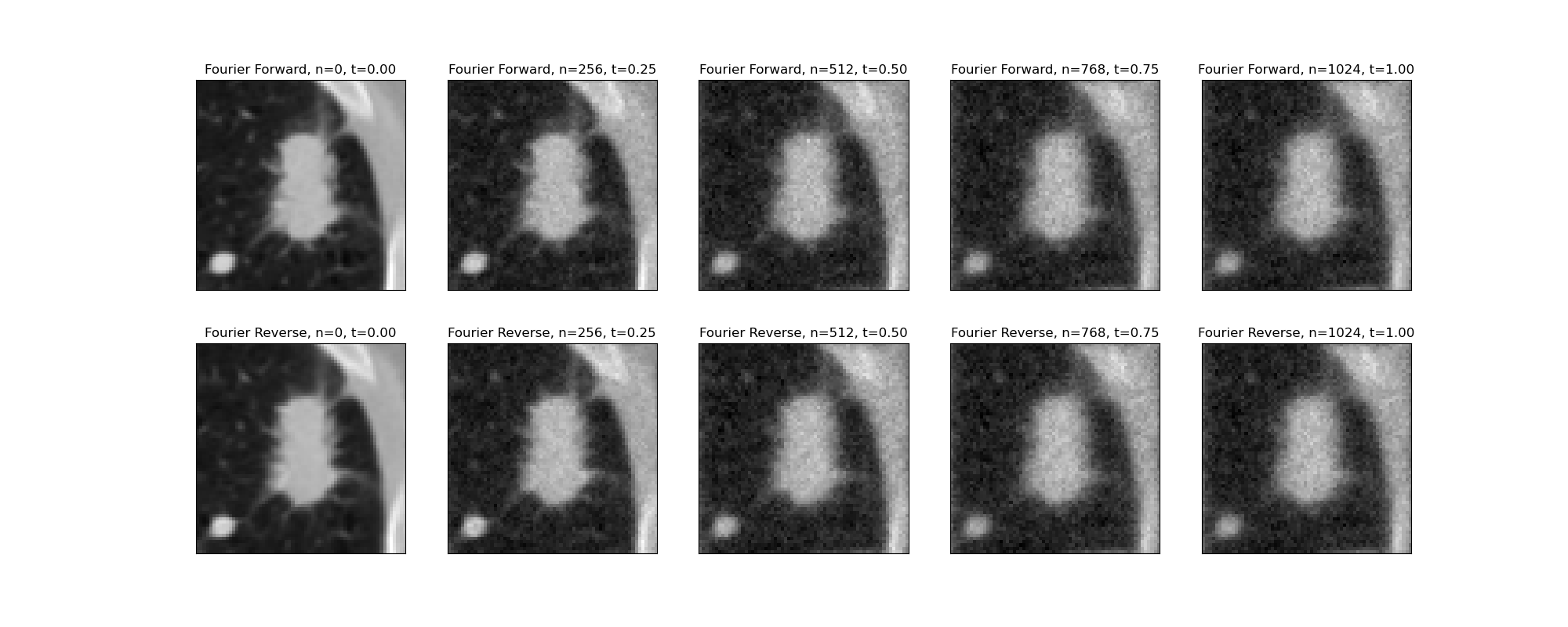}
    \caption{Forward and reverse stochastic process for the Fourier diffusion model with 1024 time steps applied to an image patch showing a lung nodule. }
    \label{fig:process_patch_fourier_1024}
\end{figure}

\begin{figure}[ht!]
    \centering
\includegraphics[trim={50mm 20mm 50mm 20mm},clip,width=0.99\textwidth]{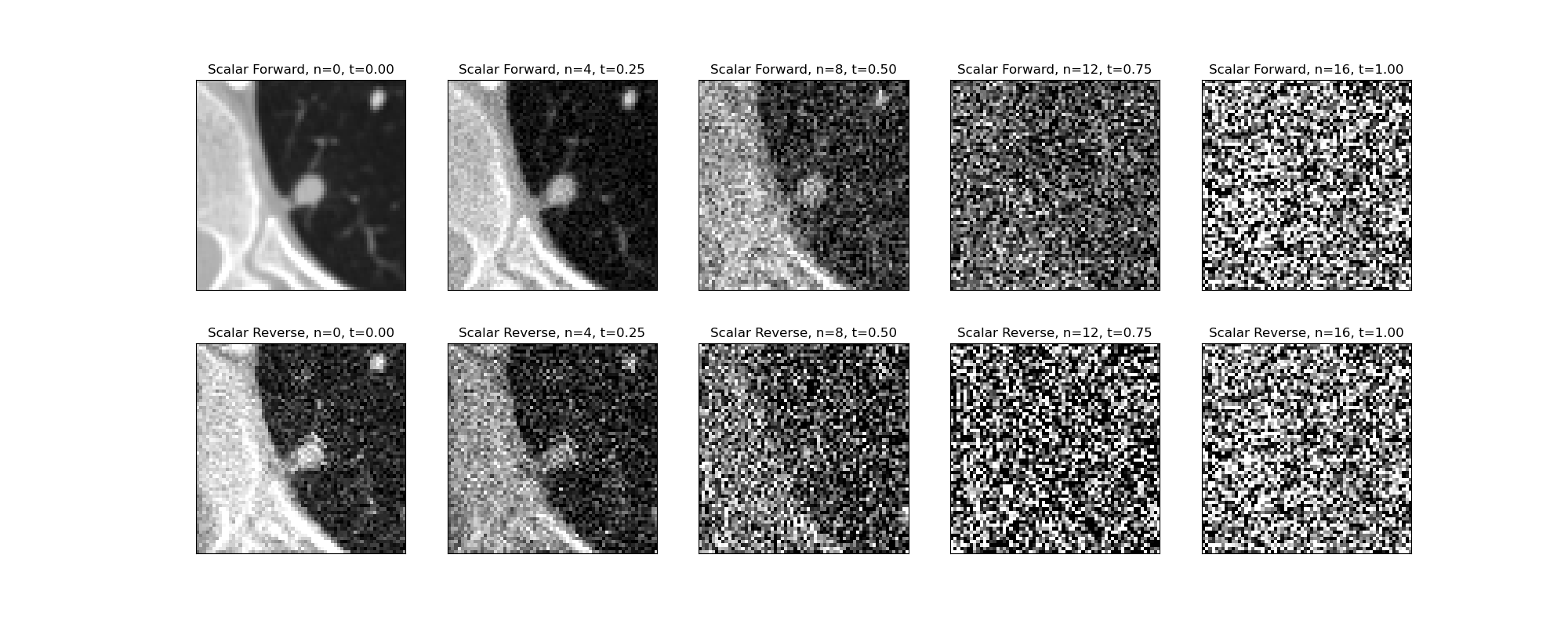}
    \caption{Forward and reverse stochastic process for the scalar diffusion model with 16 time steps applied to an image patch showing a lung nodule.  }
    \label{fig:process_patch_scalar_16}
\end{figure}

\begin{figure}[ht!]
    \centering
\includegraphics[trim={50mm 20mm 50mm 20mm},clip,width=0.99\textwidth]{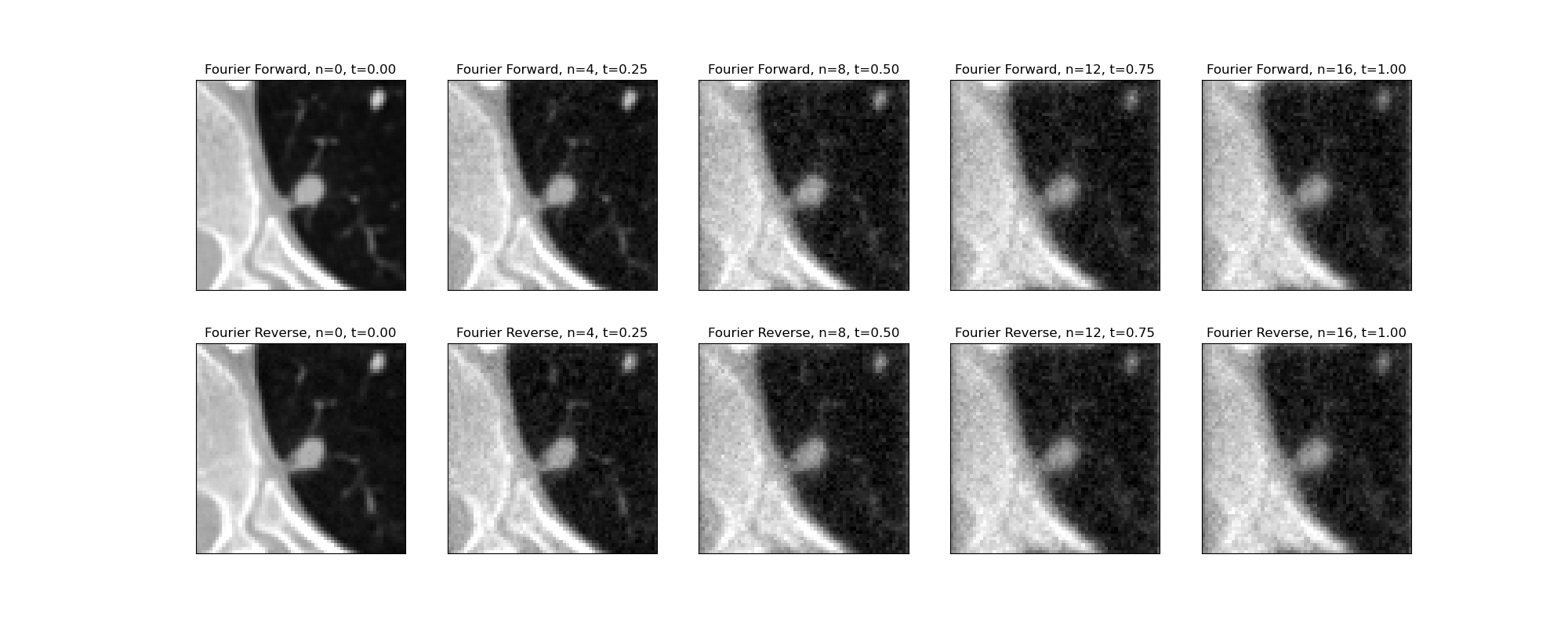}
    \caption{Forward and reverse stochastic process for the Fourier diffusion model with 16 time steps applied to an image patch showing a lung nodule. }
    \label{fig:process_patch_fourier_16}
\end{figure}

\vspace{-4mm}

\section{Results}

An example of the forward and reverse process for scalar diffusion models is displayed in Figure \ref{fig:process_scalar_1024} for a full CT image and Figure \ref{fig:process_patch_scalar_1024}  for a zoomed patch showing a lung nodule. This shows the existing method for diffusion models, which will be our reference to evaluate our new proposed method. The forward process is initialized, at $t=0$, with the ground truth images. The signal fades to zero over time in the forward process, and Gaussian white noise is added at each time step. The final result is approximately zero-mean identity-covariance Gaussian noise. The score matching neural network is trained to run the reverse process, sampling high quality images given low-radiation-dose  CT measurements. For the processes shown in Figure  \ref{fig:process_scalar_1024} and Figure \ref{fig:process_patch_scalar_1024}, we used 1024 time steps to run the reverse process. Comparing the top row and the bottom row, the samples from the reverse process appear to have similar image quality to the forward process. The final result of the reverse process at $t=0$ is a posterior estimate, or an approximation of the ground truth, given the low-radiation-dose CT measurements. Examples of the Fourier diffusion model with 1024 time steps are shown in Figure \ref{fig:process_fourier_1024} and \ref{fig:process_fourier_16}. The forward process for this case begins with the true images at $t=0$ and converges to the same distribution as the measured images at $t=1$. Note the final column of the Fourier diffusion model shows an example from the same distribution as the measured images. All the reverse processes in these Figures are for conditional image generation; so both the scalar and Fourier diffusion models are guided by measurements with the same image quality shown at $t=1$ in the Fourier diffusion models.

Figures \ref{fig:process_scalar_1024}, \ref{fig:process_patch_scalar_1024}, \ref{fig:process_fourier_1024} and \ref{fig:process_patch_fourier_1024} use 1024 time steps, which means one reverse process sample requires 1024 passes of the score-matching neural network. Corresponding examples using only 16 time steps are shown in Figures \ref{fig:process_scalar_16}, \ref{fig:process_patch_scalar_16}, \ref{fig:process_fourier_16}, and  \ref{fig:process_patch_scalar_16}, respectively. For the case of the scalar diffusion model with fewer time steps shown in Figure \ref{fig:process_scalar_16} and Figure \ref{fig:process_patch_scalar_16}, the image quality in the reverse process is much worse than the forward process. Comparing the 1024 time step reverse process, shown in Figure \ref{fig:process_scalar_1024}, with the 16 time step reverse process, shown in Figure \ref{fig:process_fourier_16}, the increased error is most likely due to time discretization. Figure \ref{fig:process_patch_fourier_16} shows an example of the Fourier diffusion model using only 16 time steps for the reverse process. Notice the improvement in image quality for the Fourier diffusion model reverse process at $t=0$ in Figure \ref{fig:process_fourier_16} relative to the Fourier diffusion model reverse process  at $t=0$ in Figure \ref{fig:process_scalar_16}. The qualitative improvement in image quality for these two cases shows a convincing visual example of improved image quality for Fourier diffusion models when using a lower number of time steps and merits further quantitative image quality analysis. 

\begin{figure}
    \centering
\includegraphics[trim={5mm 0mm 15mm 5mm},clip,width=0.32\textwidth]{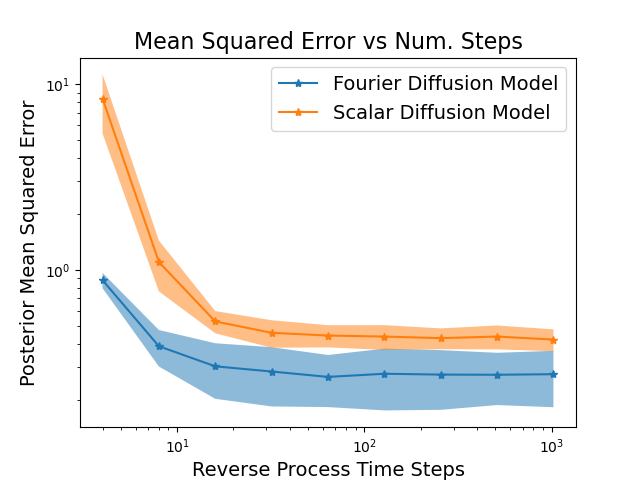}
\includegraphics[trim={3mm 0mm 15mm 5mm},clip,width=0.32\textwidth]{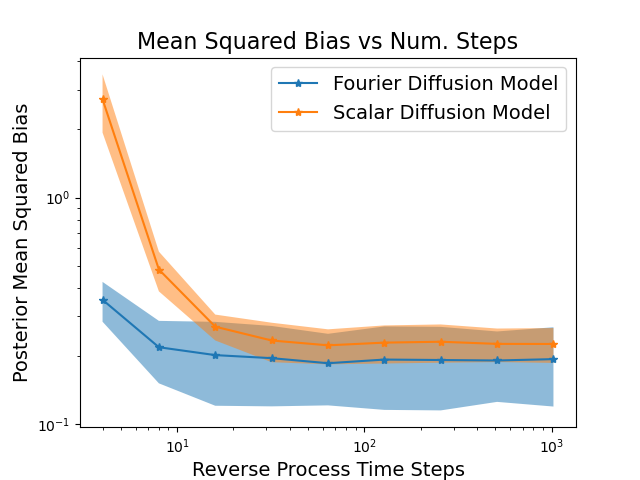}
\includegraphics[trim={3mm 0mm 15mm 5mm},clip,width=0.32\textwidth]{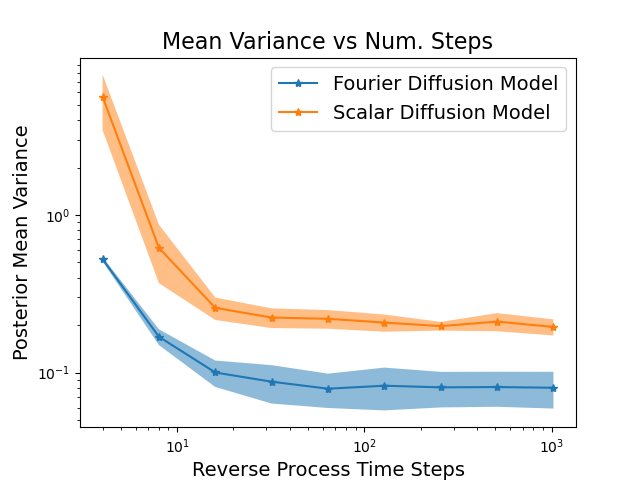}
    \caption{Mean squared bias vs number of time steps for the Fourier diffusion model and the scalar diffusion model. The shaded region shows one standard deviation of the metric over the population of validation images. }
    \label{fig:image_quality_metrics}
\end{figure}

Figure \ref{fig:image_quality_metrics} shows the mean squared error, mean squared bias, and mean variance for scalar diffusion models and Fourier diffusion models. Here, the mean refers to spatial average over the images. The line plot represents the sample mean for the population of validation images and the shaded region represents one standard deviation over the population. From these plots, we conclude that Fourier diffusion models out-perform scalar diffusion models overall. All three metrics show improved performance for the Fourier diffusion models. In particular, we note the improved performance at a low number of time steps. Fourier diffusion models with only 8 time steps achieve similar mean squared error to scalar diffusion models using 128 or even 1024 time steps. The next section provides explanations and conclusions for these results.

\vspace{-6mm}

\section{Conclusion}

\vspace{-2mm}

The results of the experiments in the previous section show that Fourier diffusion models achieve higher performance than scalar diffusion models across multiple image quality metrics and number of time steps. The improved performance may be related to the greater apparent similarity between the initial images at $t=0$ and final images at $t=1$ for Fourier diffusion models relative to scalar diffusion models. It follows that the reverse process updates for the Fourier diffusion model are smaller than those of the scalar diffusion model, which may result in improved performance for a neural network with a fixed number of parameters. Intuitively, some denoising problems are harder than others and harder denoising problems require more computational power. The neural network used for the scalar diffusion model reverse updates must dedicate some of its computational power to inverting the imagined artificial process of the image signal fading to zero; whereas the Fourier diffusion model reverse updates are completely dedicated to moving the measured image distribution towards the true image distribution. Another possible explanation is the similarity between the LSI systems of the Fourier diffusion model and the convolutional layers of the neural network. It is possible that convolutional neural networks are better suited to model local sharpening and denoising operations of the Fourier diffusion model reverse updates, as opposed to the image-wide effects in the scalar diffusion models. 

While this work was originally motivated by the goal of controlling MTF and NPS in the forward process, we note that the derivations in the earlier sections have not relied on any special properties of circulant matrices.  Therefore, we believe it should be possible to train score-based generative machine learning models defined by any Gaussian stochastic process composed of linear systems and additive Gaussian noise without specifying shift invariant systems or stationary noise. We describe the general version of the stochastic differential equations, score-matching loss function, and reverse process sampling procedure in Appendix B. In future work, we hope to explore new applications of this more general model.

Our final conclusion is that Fourier diffusion models have the potential to improve performance for conditional image generation relative to conventional scalar diffusion models. Fourier diffusion models can apply to medical imaging systems that are approximately shift invariant with stationary Gaussian noise. For the low-radiation-dose CT image restoration example, these improvements have the potential to improve image quality, diagnostic accuracy and precision, and patient health outcomes while keeping radiation dose at a suitable level for patient screening applications. We look forward to exploring new medical imaging applications of Fourier diffusion models in the future.

% \begin{figure}
%     \centering
% \includegraphics[width=0.9\textwidth]{figures/discrete_time_root_mean_squared_error.png}
%     \caption{Mean squared error vs number of time steps for the Fourier diffusion model and the scalar diffusion model. }
%     \label{fig:mean_squared_error}
% \end{figure}

% \begin{figure}
%     \centering
% \includegraphics[width=0.9\textwidth]{figures/discrete_time_root_mean_squared_bias.png}
%     \caption{Mean squared bias vs number of time steps for the Fourier diffusion model and the scalar diffusion model. }
%     \label{fig:mean_squared_bias}
% \end{figure}

% \begin{figure}
%     \centering
% \includegraphics[width=0.9\textwidth]{figures/discrete_time_root_mean_variance.png}
%     \caption{Mean squared bias vs number of time steps for the Fourier diffusion model and the scalar diffusion model. }
%     \label{fig:mean_variance}
% \end{figure}

\bibliography{main}
\bibliographystyle{spiejour}

\newpage

\section*{Appendix A}

The continuous-time stochastic process defined in \eqref{eq:x_t} is 
\begin{equation}
\mathbf{x}^{(t)} = \mathbf{H}^{(t)}_\text{MTF} \mathbf{x}^{(0)} + {\boldsymbol{\Sigma}_\text{NPS}^{(t)}}^{\hspace{-3mm}  1/2} \boldsymbol{\epsilon}^{(t)}  
\end{equation}

\noindent The change cause by a time step $\Delta t$ is:

\begin{gather}
    \mathbf{x}^{(t + \Delta t)} = \mathbf{H}^{(t + \Delta t)}_\text{MTF} {\mathbf{H}^{(t)}_\text{MTF}}^{\hspace{-3mm}-1} \mathbf{x}^{(t)} + (\boldsymbol{\Sigma}_\text{NPS}^{(t + \Delta t)} -  {\mathbf{H}^{(t+\Delta t)}_\text{MTF}}^{\hspace{-0mm} 2} {\mathbf{H}^{(t)}_\text{MTF}}^{\hspace{-3mm}-2} \boldsymbol{\Sigma}_\text{NPS}^{(t)} )^{1/2} \boldsymbol{\eta}^{(t, \Delta t)}
    \label{eq:x_t_plus_Deltat}
\end{gather}

\noindent where $\boldsymbol{\eta}^{(t, \Delta t)}$ is  zero-mean identity-covariance Gaussian noise and is independent of $\boldsymbol{\epsilon}^{(t)}$. Subtracting $\mathbf{x}^{(t)}$ yields

\begin{gather}
    \mathbf{x}^{(t + \Delta t)} - \mathbf{x}^{(t)}  =  (\mathbf{H}^{(t + \Delta t)}_\text{MTF} {\mathbf{H}^{(t)}_\text{MTF}}^{\hspace{-3mm}-1} - \mathbf{I}) \mathbf{x}^{(t)} + (\boldsymbol{\Sigma}_\text{NPS}^{(t + \Delta t)} -  {\mathbf{H}^{(t+\Delta t)}_\text{MTF}}^{\hspace{-0mm} 2} {\mathbf{H}^{(t)}_\text{MTF}}^{\hspace{-3mm}-2} \boldsymbol{\Sigma}_\text{NPS}^{(t)} )^{1/2} \boldsymbol{\eta}^{(t, \Delta t)}
    \label{eq:x_t_difference}
\end{gather}

The first term of \eqref{eq:x_t_difference} can be algebraically rearranged as follows:

\begin{gather}
    (\mathbf{H}^{(t + \Delta t)}_\text{MTF} {\mathbf{H}^{(t)}_\text{MTF}}^{\hspace{-3mm}-1} - \mathbf{I}) \mathbf{x}^{(t)} \\
    (\mathbf{H}^{(t + \Delta t)}_\text{MTF} {\mathbf{H}^{(t)}_\text{MTF}}^{\hspace{-3mm}-1} - \mathbf{H}^{(t)}_\text{MTF} {\mathbf{H}^{(t)}_\text{MTF}}^{\hspace{-3mm}-1}) \mathbf{x}^{(t)} \\
    (\mathbf{H}^{(t + \Delta t)}_\text{MTF}  - \mathbf{H}^{(t)}_\text{MTF}) {\mathbf{H}^{(t)}_\text{MTF}}^{\hspace{-3mm}-1} \mathbf{x}^{(t)} \\
    \frac{\mathbf{H}^{(t + \Delta t)}_\text{MTF}  - \mathbf{H}^{(t)}_\text{MTF}}{\Delta t} {\mathbf{H}^{(t)}_\text{MTF}}^{\hspace{-3mm}-1} \mathbf{x}^{(t)} \Delta t \label{eq:first_term}
\end{gather}

Taking the limit of \eqref{eq:first_term} as $\Delta t$ approaches zero yields

\begin{gather}
    \lim_{\Delta t \rightarrow 0} \frac{\mathbf{H}^{(t + \Delta t)}_\text{MTF}  - \mathbf{H}^{(t)}_\text{MTF}}{\Delta t} {\mathbf{H}^{(t)}_\text{MTF}}^{\hspace{-3mm}-1} \mathbf{x}^{(t)} \Delta t = \mathbf{H^{'}}^{(t)}_\text{MTF}{\mathbf{H}^{(t)}_\text{MTF}}^{\hspace{-3mm}-1} \mathbf{x}^{(t)} \text{dt}
\end{gather}

\noindent where $\mathbf{H^{'}}^{(t)}_\text{MTF} = \frac{\text{d}}{\text{dt}} \mathbf{H}^{(t)}_\text{MTF} $

The second term of \eqref{eq:x_t_difference} can also be algebraically rearranged as follows:

\begin{gather}
    (\boldsymbol{\Sigma}_\text{NPS}^{(t + \Delta t)} -  {\mathbf{H}^{(t+\Delta t)}_\text{MTF}}^{\hspace{-0mm} 2} {\mathbf{H}^{(t)}_\text{MTF}}^{\hspace{-3mm}-2} \boldsymbol{\Sigma}_\text{NPS}^{(t)} )^{1/2} \boldsymbol{\eta}^{(t, \Delta t)}  \\
    (\boldsymbol{\Sigma}_\text{NPS}^{(t + \Delta t)} - ( {\mathbf{H}^{(t+\Delta t)}_\text{MTF}} {\mathbf{H}^{(t)}_\text{MTF}}^{\hspace{-3mm}-1})^2 \boldsymbol{\Sigma}_\text{NPS}^{(t)} )^{1/2} \boldsymbol{\eta}^{(t, \Delta t)}  \\
    (\boldsymbol{\Sigma}_\text{NPS}^{(t + \Delta t)} - (\mathbf{I} +  {\mathbf{H}^{(t+\Delta t)}_\text{MTF}}{\mathbf{H}^{(t)}_\text{MTF}}^{\hspace{-3mm}-1} - \mathbf{I})^{2} \boldsymbol{\Sigma}_\text{NPS}^{(t)} )^{1/2} \boldsymbol{\eta}^{(t, \Delta t)}  \\
    (\boldsymbol{\Sigma}_\text{NPS}^{(t + \Delta t)} - (\mathbf{I} +  {\mathbf{H}^{(t+\Delta t)}_\text{MTF}}{\mathbf{H}^{(t)}_\text{MTF}}^{\hspace{-3mm}-1} -  {\mathbf{H}^{(t)}_\text{MTF}}{\mathbf{H}^{(t)}_\text{MTF}}^{\hspace{-3mm}-1})^{2} \boldsymbol{\Sigma}_\text{NPS}^{(t)} )^{1/2} \boldsymbol{\eta}^{(t, \Delta t)}  \\
    (\boldsymbol{\Sigma}_\text{NPS}^{(t + \Delta t)} - (\mathbf{I} +  ({\mathbf{H}^{(t+\Delta t)}_\text{MTF}} -  {\mathbf{H}^{(t)}_\text{MTF}}) {\mathbf{H}^{(t)}_\text{MTF}}^{\hspace{-3mm}-1})^{2} \boldsymbol{\Sigma}_\text{NPS}^{(t)} )^{1/2} \boldsymbol{\eta}^{(t, \Delta t)}  \\
    (\boldsymbol{\Sigma}_\text{NPS}^{(t + \Delta t)} - (\mathbf{I} +  \frac{{\mathbf{H}^{(t+\Delta t)}_\text{MTF}} -  {\mathbf{H}^{(t)}_\text{MTF}}}{\Delta t} {\mathbf{H}^{(t)}_\text{MTF}}^{\hspace{-3mm}-1} \Delta t)^{2} \boldsymbol{\Sigma}_\text{NPS}^{(t)} )^{1/2} 
    \boldsymbol{\eta}^{(t, \Delta t)} \\
    (\frac{\boldsymbol{\Sigma}_\text{NPS}^{(t + \Delta t)} - (\mathbf{I} + 
    2\frac{\mathbf{H}^{(t+\Delta t)}_\text{MTF} -  {\mathbf{H}^{(t)}_\text{MTF}}}{\Delta t} {\mathbf{H}^{(t)}_\text{MTF}}^{\hspace{-3mm}-1} \Delta t + \mathcal{O}(\Delta t^2)) \boldsymbol{\Sigma}_\text{NPS}^{(t)}}{\Delta t} )^{1/2}
     \sqrt{\Delta t} \enspace \boldsymbol{\eta}^{(t, \Delta t)} \\
    (-  2 \frac{{\mathbf{H}^{(t+\Delta t)}_\text{MTF}} -  {\mathbf{H}^{(t)}_\text{MTF}}}{\Delta t} {\mathbf{H}^{(t)}_\text{MTF}}^{\hspace{-3mm}-1}  \boldsymbol{\Sigma}_\text{NPS}^{(t)} + \frac{\boldsymbol{\Sigma}_\text{NPS}^{(t + \Delta t)} - \boldsymbol{\Sigma}_\text{NPS}^{(t)}}{\Delta t}  - \frac{\mathcal{O}(\Delta t^2)}{\Delta t} \boldsymbol{\Sigma}_\text{NPS}^{(t)}  )^{1/2} \sqrt{\Delta t} \enspace \boldsymbol{\eta}^{(t, \Delta t)} \label{eq:second_term}
\end{gather}

\noindent where $\mathcal{O}(\Delta t^2)$ indicates second order and higher terms of the Taylor expansion. Note, we have made the approximation that $\frac{\mathbf{H}^{(t+\Delta t)}_\text{MTF} -  {\mathbf{H}^{(t)}_\text{MTF}}}{\Delta t}$ can be considered as a constant with respect to $\Delta t$ for the purposes of the Taylor expansion, which is valid for continuously differentiable functions of time in the limit of small values of $\Delta t$. Taking the limit of \eqref{eq:second_term} as $\Delta t$ approaches zero yields

\begin{gather}
    \lim_{\Delta t \rightarrow 0} (-  2 \frac{{\mathbf{H}^{(t+\Delta t)}_\text{MTF}} -  {\mathbf{H}^{(t)}_\text{MTF}}}{\Delta t} {\mathbf{H}^{(t)}_\text{MTF}}^{\hspace{-3mm}-1}  \boldsymbol{\Sigma}_\text{NPS}^{(t)} + \frac{\boldsymbol{\Sigma}_\text{NPS}^{(t + \Delta t)} - \boldsymbol{\Sigma}_\text{NPS}^{(t)}}{\Delta t} + \frac{\mathcal{O}(\Delta t^2)}{\Delta t} \boldsymbol{\Sigma}_\text{NPS}^{(t)})^{1/2} \sqrt{\Delta t} \enspace \boldsymbol{\eta}^{(t, \Delta t)} \\
    (-  2 {\mathbf{H^{'}}^{(t)}_\text{MTF}} {\mathbf{H}^{(t)}_\text{MTF}}^{\hspace{-3mm}-1}  \boldsymbol{\Sigma}_\text{NPS}^{(t)} + \boldsymbol{\Sigma^{'}}_\text{NPS}^{(t)}  )^{1/2} \mathbf{dw}
\end{gather}

\noindent where $\boldsymbol{\Sigma^{'}}_\text{NPS}^{(t)} = \frac{\text{d}}{\text{dt}} \boldsymbol{\Sigma}_\text{NPS}$ and $\mathbf{dw}$ is infinitesimal white Gaussian noise with covariance, $\text{dt} \mathbf{I}$. Therefore, the limit of \eqref{eq:x_t_difference} as $\Delta t$ approaches zero is:

\begin{equation}
    \mathbf{dx} = \mathbf{H^{'}}^{(t)}_\text{MTF}{\mathbf{H}^{(t)}_\text{MTF}}^{\hspace{-3mm}-1} \mathbf{x}^{(t)} \text{dt} + (-  2 {\mathbf{H^{'}}^{(t)}_\text{MTF}} {\mathbf{H}^{(t)}_\text{MTF}}^{\hspace{-3mm}-1}  \boldsymbol{\Sigma}_\text{NPS}^{(t)} + \boldsymbol{\Sigma^{'}}_\text{NPS}^{(t)}  )^{1/2} \mathbf{dw}
\end{equation}

\newpage

\section*{Appendix B}

This work was motivated by the goal of controlling spatial resolution and noise in the spatial frequency domain. However, we have not relied on any special properties of circulant matrices. Therefore, we can write a more general version that may be useful for some cases. Consider a forward stochastic process where update steps are defined by the linear system, $\mathbf{H}^{(t)}$ and additive Gaussian noise with covariance, $\boldsymbol{\Sigma}^{(t)}$ as defined below:

\begin{gather}
\mathbf{x}^{(t)} = \mathbf{H}^{(t)} \mathbf{x}^{(0)} + {\boldsymbol{\Sigma}^{(t)}}^{\hspace{0mm}  1/2} \boldsymbol{\epsilon}^{(t)}  .
\label{eq:x_t_general}
\end{gather}

\noindent The corresponding forward stochastic differential equation is

\begin{equation}
    \mathbf{dx} = \mathbf{H^{'}}^{(t)}{\mathbf{H}^{(t)}}^{\hspace{0mm}-1} \mathbf{x}^{(t)} \text{dt} + (-  2 {\mathbf{H^{'}}^{(t)}} {\mathbf{H}^{(t)}}^{\hspace{0mm}-1}  \boldsymbol{\Sigma}^{(t)} + \boldsymbol{\Sigma^{'}}^{(t)}  )^{1/2} \mathbf{dw} , \label{eq:SDE_general}
\end{equation}

\noindent and the time-reversed stochastic differential equation is

\begin{gather}
     \mathbf{dx} = [\mathbf{H^{'}}^{(t)}{\mathbf{H}^{(t)}}^{\hspace{0mm}-1} \mathbf{x}^{(t)} - (-  2 {\mathbf{H^{'}}}^{(t)} {{\mathbf{H}}^{(t)}}^{\hspace{0mm}-1}  
     \boldsymbol{\Sigma}^{(t)} + \boldsymbol{\Sigma^{'}}^{(t)} )  \nabla_{\mathbf{x}^{(t)}} \log{\text{p} (\mathbf{x}^{(t)}|\mathbf{y})}] \text{dt} \nonumber \\
     \hspace{70mm} + (-  2 {\mathbf{H^{'}}}^{(t)} {\mathbf{H}^{(t)}}^{\hspace{0mm}-1}  \boldsymbol{\Sigma}^{(t)} + \boldsymbol{\Sigma^{'}}^{(t)}  )^{1/2} \mathbf{dw}.
\end{gather}

\noindent For this general case, the score-matching loss function is

\begin{equation}
\underset{\mathbf{x}^{(0)}, t}{\mathbb{E}}[(\boldsymbol{\hat{\epsilon}}_{\boldsymbol{\theta}}(\mathbf{x}^{(t)}, \mathbf{y}, t) - \boldsymbol{\epsilon}^{(t)})^T {\boldsymbol{\Sigma}^{(t)}}^{\hspace{0mm}-1} (\boldsymbol{\hat{\epsilon}}_{\boldsymbol{\theta}}(\mathbf{x}^{(t)}, \mathbf{y}, t) - \boldsymbol{\epsilon}^{(t)}) ]
\end{equation}

For the experiments in this article, we focus on the forward process in \eqref{eq:x_t} and score-matching loss function in \eqref{eq:score_matching_loss} using circulant matrices to train measurement-conditioned Fourier diffusion models for medical image restoration. In the future, we are interested in exploring more applications of the more general equations above.

\end{document}